\documentclass[sn-mathphys,Numbered,iicol]{sn-jnl}
\usepackage{graphicx}
\usepackage{multirow}
\usepackage{amsmath,amssymb,amsfonts}
\usepackage{amsthm}
\usepackage{tikz}
\usepackage[export]{adjustbox}
\usepackage{pgfplots}
\pgfplotsset{width=6.8cm,compat=newest}
\usepackage{subcaption}
\usepackage[labelfont=bf]{caption}
\usetikzlibrary{positioning}
\usepackage{changepage}
\usepackage{mathrsfs}
\usepackage[title]{appendix}
\usepackage{xcolor}
\usepackage{textcomp}
\usepackage{manyfoot}
\usepackage{booktabs}
\usepackage{algorithm}
\usepackage{algorithmicx}
\usepackage{algpseudocode}
\usepackage{listings}
\theoremstyle{thmstyleone}

\theoremstyle{thmstyletwo}

\theoremstyle{thmstylethree}

\begin{document}

\title[Higher-dimensional Holographic Superconductors in Born-Infeld Electrodynamics and $f(R)$ Gravity]{Higher-dimensional Holographic Superconductors in Born-Infeld Electrodynamics and $f(R)$ Gravity}
\author*{\fnm{Alexandar} \sur{Roussev}}\email{alexandar.roussev@stud.uni-heidelberg.de}
\affil{\orgname{Heidelberg University}, \orgaddress{\street{Grabengasse 1}, \city{69117 Heidelberg}, \country{Germany}}}

\abstract{In this paper, the properties of higher dimensional holographic superconductors are studied in the background of $f(R)$ gravity and Born-Infeld electrodynamics. A specific model of $f(R)$ gravity is considered, allowing a perturbative approach to the problem. The Sturm-Liouville eigenvalue problem is used to analytically calculate the critical temperature and the condensation operator. An expression for the critical temperature in terms of the charge density including the correction from modified gravity is derived. It is seen that the higher values of the Born-Infeld coupling parameter make the condensation harder to form. In addition, the limiting values of this parameter, above which Born-Infeld electrodynamics cannot be applied, are found for different dimensions. Another interesting property is that the increasing modifications of $f(R)$ gravity lead to larger values of the critical temperature and a decrease in the condensation gap, which means that the condensation is easier to form.}

\maketitle

\section{\label{sec1}Introduction}
The discovery of the AdS/CFT correspondence \cite{Maldacena_1999,1998AdTMP...2..253W}, which states that a $d$-dimensional asymptotically AdS spacetime can be described by a $(d-1)$-dimensional conformal field theory on the boundary, has allowed theorists to use a gravitational description of problems in condensed matter physics. One such application in superconductivity has been researched for the past 15 years. For the first time, in \cite{2005CQGra..22.5121G,2008PhRvD..78f5034G,2008PhRvL.101c1601H} it was shown that the gauge/gravity duality can allow the gravitational description of superconductor phase transitions with the help of black holes.\\
There are numerous models of such holographic superconductors that have been researched since then. These include superconductors in different dimensions \cite{2018EPJC...78..984M,2010PhRvD..81j6007P,2008PhRvD..78l6008H, 2016EPJC...76..146G}, as well as holographic superconductors in different backgrounds, e.g. in an external magnetic field \cite{2008JHEP...09..121A, 2009arXiv0906.0519A, 2010JHEP...08..108G, 2012PhRvD..86j6009R,2014MPLA...2950088G, 2019JHEP...11..085C}, with Weyl corrections \cite{2011PhLB..697..153W, 2016IJGMM..1350131M,2020EPJC...80.1059L}, in Horava-Lifshitz gravity \cite{2010PhRvD..81f6003C,2011JHEP...05..118M,2015IJMPD..2450038L,2016PhLB..759..184L}, Gauss-Bonnet gravity \cite{2009JHEP...10..010G,2011JHEP...04..028L,2010JHEP...08..108G,2010PhRvD..81j6007P,2010PhLB..693..159P,2010JHEP...12..029B,2011JHEP...11..088P,2011JPhCS.283a2016G,2011JHEP...10..044B,2011PhRvD..83f6010J,2012JHEP...05..156G,2013JHEP...05..101Y,2012arXiv1202.3586L,2016EPJC...76..146G, Lu:2016smd, 2020JHEP...12..192Q, 2022EPJP..137..852M}, and $f(R)$ gravity \cite{2013arXiv1306.2082X,2014EPJP..129...30M}. Some studies also include non-linear electrodynamics \cite{2014EPJP..129...30M,2012PhRvD..86j6009R,2010PhLB..686...68J,2011JHEP...11..045J,2012JHEP...05..002G,2011PhRvD..83f6010J,2012JHEP...05..156G,2012arXiv1212.2721B, 2012GReGr..44.1309W,2014MPLA...2950088G,2020EPJC...80..928M,2020PhRvD.101b6012H}. In particular, Born-Infeld electrodynamics is especially interesting: it has finite self energies for charged point particles, and is the only non-linear electromagnetic theory that possesses invariance under electromagnetic duality and has no birefringence. This theory has also had important applications in cosmology \cite{2000IJMPA..15.4341G,2003CQGra..20.5425G,2002JHEP...09..026F,2010MPLA...25...35M,PhysRevD.66.103501,P Vargas Moniz_2002,2010CQGra..27w5009V,2011PhRvD..83f4014H}, such as predicting early-time inflation. Furthermore, the Born-Infeld action can describe D-branes at low energy in string theory \cite{2000mfsw.book..417T,1997NuPhB.501...41T} - one of the main reasons for the renewed interest in Born-Infeld theory. \\
We can see that some modified gravity theories, such as Gauss-Bonnet gravity, have been researched thoroughly. However, that is not the case for generic $f(R)$ gravity. This framework has very special properties, making it an essential part of the study of modified gravity theories. For example, $f(R)$ gravity can give an alternative explanation to cosmological phenomena that are otherwise related to the introduction of exotic dark components, such as dark energy and dark matter. This higher-order theory predicts both early-time inflation and late-time cosmic acceleration, and is also the only modified gravity theory that avoids the Ostrogradsky instability \cite{2007LNP...720..403W, 2008PhRvD..78l4007H}. $f(R)$ gravity is known to be Lorentz invariant, while higher-order curvature corrections appear in string theory as well, which makes $f(R)$ gravity suitable for models using AdS/CFT \cite{2013arXiv1306.2082X, 2014EPJP..129...30M, 2010RvMP...82..451S}. Thus, $f(R)$ gravity is an appropriate framework for the analysis of holographic superconductors.\\ 
An additional inspiration to consider the combination of $f(R)$ gravity and Born-Infeld electrodynamics is the fact that they have been particularly important for higher-dimensional brane-world models (see \cite{2010LRR....13....3D, 2013PhRvD..87j4001B, 2000JHEP...09..013G} and references therein). Hence, we note that these two theories have a key role in both cosmology and string theory. \\
Currently, holographic superconductors in $f(R)$ gravity have been analyzed in 4-dimensional spacetime in linear and nonlinear Maxwell electrodynamics only. Motivated by all the aforementioned features of $f(R)$ gravity and Born-Infeld electrodynamics, this paper presents analysis of holographic superconductors in the context of both of these theories, while also giving a generalization to higher dimensions, which allows the applicability of the results here in the investigation of higher-dimensional theories (e.g. string theory). \\
One common method in analyzing holographic superconductor models is the matching method \cite{2009JHEP...10..010G,2010PhRvD..81j6007P,2013arXiv1306.2082X, 2014MPLA...2950088G,2013JHEP...05..101Y, 2014EPJP..129...30M, 2011JHEP...05..118M}. The idea behind this analytic method is to take the solutions of the field equations near the horizon and the boundary and to match them at an intermediate point. Another analytic method is based on the Sturm-Liouville (SL) eigenvalue problem \cite{2012arXiv1212.2721B,2016EPJC...76..146G, 2016PhLB..759..184L, 2018EPJC...78..984M, 2011JHEP...04..028L, 2012JHEP...05..156G,2011JHEP...11..045J,2012JHEP...05..002G,2018PhLB..781..139S}, which has been shown to yield more accurate results \cite{2014MPLA...2950088G, 2012JHEP...08..104G,2011JHEP...11..088P}. For this reason, this paper will employ the SL method in the following analysis. \\
In this paper, we will look at higher dimensional holographic superconductors in Born-Infeld electrodynamics and $f(R)$ gravity in the probe limit using the SL method. The framework of $f(R)$ gravity is analyzed using perturbative techniques. We get the equation relating the critical temperature and the charge density in $d$-dimensions, while also studying the cases for $d=5,6,7$ more thoroughly. It is seen that the critical temperature decreases with the larger values of the Born-Infeld parameter, as expected. However, it increases with larger modifications of $f(R)$ gravity, while the condensation gap becomes smaller. This shows that the condensation is easier to form when the $f(R)$ gravity configuration considered here is present. \\
The structure of the paper is as follows. In Section \ref{sec2} the particular model of $f(R)$ gravity is constructed. In Section \ref{sec3} the basic setup of the holographic superconductors is given. In Section \ref{sec4} and Section \ref{sec5} the critical temperature and the condensate, respectively, are computed. Section \ref{sec6} contains concluding remarks.
\section{Model of $f(R)$ gravity}
\label{sec2}
In this section, the specific model of $f(R)$ gravity that we are going to study in this paper is introduced. We will follow some of the steps in \cite{2009EPJC...64..113C,2013arXiv1306.2082X, 2016EPJC...76..146G, 2006PhRvD..74f4022M, 2012MPLA...2750138S, 2010MPLA...25.1281S}. \\
First, we begin the analysis by writing down the action for $f(R)$ gravity with a matter field in $d$-dimensions:
\begin{eqnarray}
\label{eq1}
S=\frac{1}{16\pi G_d}\int d^d x \sqrt{-g}\left(f(R)+16\pi G_d \mathcal{L}_m\right),
\end{eqnarray}
where $f(R)$ is a function of the Ricci scalar and $G_d$ is the $d$-dimensional Newtonian gravitational constant. We consider the probe limit, so we can take $G_d \to 0$. This is equivalent to considering the action in the absence of matter fields. Therefore, we have for the following equation of motion:
\begin{eqnarray}
\label{eq2}
    \resizebox{0.43\textwidth}{!}{$R_{\mu \nu} F(R)-\frac{1}{2}f(R)g_{\mu \nu}+(g_{\mu \nu} \nabla^2-\nabla_{\mu} \nabla_{\nu})F(R)=0,$}
\end{eqnarray}
where $F(R)=\frac{df(R)}{dR}$. Taking the trace, we obtain:
\begin{eqnarray}
\label{eq3}
    F(R)R-\frac{d}{2}f(R)+(d-1)\nabla^2F(R)=0.
\end{eqnarray}
Therefore, we have for $f(R)$
\begin{eqnarray}
\label{eq4}
    f(R)=\frac{2}{d}(F(R)R+(d-1)\nabla^2F(R)).
\end{eqnarray}
After substituting this in Eq.(\ref{eq2}), we get
\begin{eqnarray}
\label{eq5}
    \resizebox{0.43\textwidth}{!}{$R_{\mu \nu}F(R)-\nabla_{\mu} \nabla_{\nu} F(R)=\frac{g_{\mu \nu}}{d}\left(F(R)R-\nabla^2F(R)\right).$}
\end{eqnarray}
Therefore, we can easily see that the expression
\begin{eqnarray}
\label{eq6}
    \frac{F(R)R_{\mu \mu}-\nabla_{\mu}\nabla_{\mu}F(R)}{g_{\mu \mu}}
\end{eqnarray}
does not depend on the index $\mu$. In order to have a superconducting phase transition, we need a planar black hole solution. For this reason, we will take the following plane-symmetric metric:
\begin{eqnarray}
\label{eq7}
    ds^2=-A(r)dt^2+B(r)dr^2+r^2h_{ij}dx^i dx^j,
\end{eqnarray}
where $h_{ij}dx^i dx^j$ is the line element of a $(d-2)$-dimensional hypersurface with zero curvature.
The only non-zero components of the Ricci tensor of this metric are
\begin{eqnarray}
\label{eq8}
    &&\resizebox{0.42\textwidth}{!}{$R_{00}=\frac{A''}{2B}-\frac{A'B'}{4B^2}-\frac{A'^2}{4AB}+\frac{A'}{2Br}(d-2);$}\nonumber \\
    &&\resizebox{0.42\textwidth}{!}{$R_{11}=-\frac{A''}{2A}+\frac{A'^2}{4A^2}+\frac{A'B'}{4AB}+\frac{B'}{2Br}(d-2);$} \\
    &&\resizebox{0.45\textwidth}{!}{$R_{22}=...=R_{ii}=...=R_{dd}=-\frac{A'r}{2AB}+\frac{B'r}{2B^2}-\frac{d-3}{B},$} \nonumber
\end{eqnarray}
where $2 \le i\le d$ and the prime denotes the derivative with respect to $r$. Thus, if we define $X=AB$, and using the fact that the quantity in Eq.(\ref{eq6}) is the same for any value of $\mu$, we have the following differential equations
\begin{eqnarray}
    \label{eq9}
    &&F''-\frac{F'}{2}\frac{X'}{X}-\frac{F(d-2)}{2r}\frac{X'}{X}=0;\\
    \label{eq10}
    &&A''+\frac{F'}{F}\left(A'-\frac{2A}{r}\right)-\frac{2A(d-3)}{r^2} \nonumber \\
    &&+\frac{A}{r}\left(\frac{B'}{B}+\frac{A'}{A}(d-3)\right)-\frac{A'}{2}\frac{X'}{X}=0.
\end{eqnarray}
Next, we will assume that $F(r)$ is of the form
\begin{eqnarray}
    \label{eq11}
    F(r)=ar+b.
\end{eqnarray}
From Eq.(\ref{eq9}) it follows that $X$ is a constant, i.e. we will take $X=c_3$. Then, solving Eq.(\ref{eq10}) gives
\begin{strip}
\begin{eqnarray}
\label{eq12}
    A(r)=c_2 r^2\left((-1)^d\frac{a^{d-1}}{b^d}\ln \left(\frac{ar+b}{r}\right)+\sum_{k=2}^{d}\frac{(-1)^{d-k}a^{d-k}}{b^{d-k+1}r^{k-1}(1-k)} \right)+c_1 r^2.
\end{eqnarray}
\end{strip}
In order for this solution to be correct in the limit of higher dimensional Schwarzschild-AdS/dS black holes, one can see that
\begin{eqnarray}
\label{eq13}
    c_2=2Mb(d-1),
\end{eqnarray}
where $M$ is the mass of the black hole. From now on, we will also take $c_3=1$ for simplicity, as was done in \cite{2013arXiv1306.2082X}.\\
Calculating the Ricci scalar, we obtain:
\begin{eqnarray}
\label{eq14}
    R(r)=&&-\frac{A''}{AB}+\frac{A'B'}{2AB^2}+\frac{A'^2}{2A^2B}-\frac{A'(d-2)}{ABr} \nonumber \\
    &&+\frac{B'(d-2)}{B^2r}-\frac{(d-2)(d-3)}{Br^2}.
\end{eqnarray}
Using Eq.(\ref{eq12}) and the fact that now $X=1$, after taking the limit $r\to \infty$ we get:
\begin{eqnarray}
\label{eq15}
    \resizebox{0.42\textwidth}{!}{$R(r)=-d(d-1)c_1-(-1)^d\frac{2d(d-1)^2a^{d-1}M}{b^{d-1}}\ln{a}.$}
\end{eqnarray}
Thus, we can see that the spacetime is asymptotically AdS/dS. For the purpose of this paper, we will take it to be AdS. Therefore, taking into account that under these conditions the Ricci scalar in Eq.(\ref{eq15}) is
\begin{eqnarray}
\label{eq16}
    R(r)=\frac{2d\Lambda}{d-2}=-\frac{d(d-1)}{l^2},
\end{eqnarray}
where $\Lambda$ is the cosmological constant and $l$ is the AdS radius, we have for the effective values of these two parameters:
\begin{eqnarray}
    \label{eq17}
    &&\Lambda_{eff}=-\frac{(d-1)(d-2)}{2} c_1 \nonumber \\
    &&-(-1)^{d}\frac{(d-1)^2(d-2)a^{d-1}M}{b^{d-1}} \ln{a}; \\
    \label{eq18}
    &&l_{eff}=\sqrt{-\frac{(d-1)(d-2)}{2\Lambda_{eff}}}.
\end{eqnarray}
For the model in this paper, the value $a$ defined in Eq.(\ref{eq11}) will be considered as a small parameter. Thus, we can calculate the Ricci scalar to the first order of $a$:
\begin{eqnarray}
\label{eq19}
    R(r) \approx d(1-d)c_1+\frac{4M(d-1)a}{(2-d)br^{d-2}}.
\end{eqnarray}
\\
From this expression we can easily obtain $r$ as a function of $R$:
\begin{eqnarray}
    \label{eq20}
    r(R)=\left(\frac{4M(d-1)a}{b(2-d)(R+d(d-1)c_1)}\right)^{\frac{1}{d-2}}.
\end{eqnarray}
We substitute this result in Eq.(\ref{eq11}) and after integration:
\begin{strip}
\begin{eqnarray}
\label{eq21}
    f(R)=c_4+bR+\frac{a(d-2)}{d-3}\left(-\frac{4aM(d-1)}{b(d-2)(c_1(d-1)d+R)}\right)^{\frac{1}{d-2}}(c_1(d-1)d+R).
\end{eqnarray}
\end{strip}
In order for this expression to be correct in the limit of $a \to 0$ (Einstein-Hilbert action), we set $b=1$ and $c_4=-2\Lambda$. If we also look at Eq.(\ref{eq15}) for $a \to 0$, using Eq.(\ref{eq16}) we see that
\begin{eqnarray}
\label{eq22}
    c_1=-\frac{2\Lambda}{(d-1)(d-2)}=\frac{1}{l^2}.
\end{eqnarray}
This is the $f(R)$ gravity model that we are going to analyze in this paper. The final form of the metric is
\begin{eqnarray}
\label{eq23}
    &&A(r)=\frac{r^2}{l^2}+2M(d-1)\bigg((-1)^d a^{d-1} r^2 \ln \left(\frac{1+ar}{r}\right) \nonumber \\
    &&+\sum_{k=2}^{d}\frac{(-1)^{d-k}a^{d-k}}{r^{k-3}(1-k)}\bigg); \\
    \label{eq24}
    &&B(r)=\frac{1}{A(r)}.
\end{eqnarray}
To find the radius of the outer horizon, we can approximate it to the first order of $a$ in the following form:
\begin{eqnarray}
\label{eq25}
    r_{+}=r_0+ar_1+\mathcal{O}(a^2).
\end{eqnarray}
Using the condition $A(r_{+})=0$, we obtain:
\begin{eqnarray}
\label{eq26}
    r_{+}=r_0-\frac{ar_0^2}{d-2},
\end{eqnarray}
where $r_0=(2Ml^2)^{\frac{1}{d-1}}$.\\
Finally, we can get the result for the Hawking temperature to the first order of $a$:
\begin{eqnarray}
    \label{eq27}
    T=\frac{A'(r_{+})}{4\pi}=\frac{(d-1)r_0}{4\pi l^2},
\end{eqnarray}
which shows that there are no corrections to the first order of $a$, agreeing with \cite{2013arXiv1306.2082X}.\\
For the rest of the paper, we will focus on the smallness of the dimensionless quantity $ar_0$ rather than just $a$, which does not change the results presented above.
\section{\label{sec3}Basic setup}
We can now start analyzing the main model. The equations here have also been derived in \cite{2016EPJC...76..146G} for a general metric.\\
The Lagrangian density $\mathcal{L}_m$ in Eq.(\ref{eq1}) can be written down as
\begin{eqnarray}
    \label{eq28}
    \mathcal{L}_m=&&\frac{1}{b}\left(1-\sqrt{1+\frac{bF_{\mu \nu}F^{\mu \nu}}{2}}\right)-\lvert\partial_{\mu}\psi -iqA_{\mu}\psi\rvert^2 \nonumber \\
    &&-m^2\lvert \psi \rvert^2,
\end{eqnarray}
where $F_{\mu \nu}=\partial_{\mu} A_{\nu}-\partial_{\nu} A_{\mu}$, and $A_{\mu}$ and $\psi$ are respectively the gauge and scalar fields. We will take the following ansatz:
\begin{eqnarray}
    \label{eq29}
    A_{\mu}dx^{\mu}=\phi(r)dt;\nonumber \\
    \psi=\psi(r).
\end{eqnarray}
Therefore, we have for the equations of motion:
\begin{eqnarray}
\label{eq30}
    &&\phi''(r)+\phi'(r)(1-b\phi'(r)^2)\frac{d-2}{r} \nonumber \\
    &&-\frac{2q^2\phi(r)\psi^2(r)}{A(r)}(1-b\phi'(r)^2)^{\frac{3}{2}}=0 \\
    \label{eq31}
    &&\psi''(r)+\psi'(r)\left(\frac{d-2}{r}+\frac{A'(r)}{A(r)}\right) \nonumber \\
    &&+\left(\frac{q^2\phi^2(r)}{A(r)}-m^2\right)\frac{\psi(r)}{A(r)}=0.
\end{eqnarray}
Due to the scaling symmetry we can also choose $q=1$ without loss of generality.\\
To solve these equations, we must first look at the boundary conditions. At the horizon, in order for the fields to be finite we require that $\phi(r_{+})=0$ and $\psi(r_{+})$ is finite. At $r \to \infty$,  for asymptotically AdS spacetime, we have:
\begin{eqnarray}
    \label{eq32}
    \phi(r)=\mu - \frac{\rho}{r^{d-3}};\\
    \label{eq33}
    \psi(r)=\frac{\psi_{-}}{r^{\Delta_{-}}}+\frac{\psi_{+}}{r^{\Delta_{+}}},
\end{eqnarray}
where
\begin{eqnarray}
    \label{eq34}
    \Delta_{\pm}=\frac{(d-1) \pm \sqrt{(d-1)^2+4m^2l^2}}{2},
\end{eqnarray}
and $\mu$ and $\rho$ are the chemical potential and the charge density in the dual field theory. We can also choose either $\psi_{-}$ or $\psi_{+}$ to vanish \cite{2008PhRvL.101c1601H}. In this paper we will have the condition $\psi_{-}=0$, while $\psi_{+}$ is dual to the expectation value of the condensation operator on the boundary. \\
If we transform the coordinates as $z=\frac{r_{+}}{r}$, Eqs.(\ref{eq30}) and (\ref{eq31}) become
\begin{eqnarray}
    \label{eq35}
    &&\phi''(z)-\frac{\phi'(z)}{z}(d-4)+\frac{(d-2)bz^3\phi'(z)^3}{r_{+}^2} \nonumber \\
    &&-\frac{2r_{+}^2\phi(z)\psi^2(z)}{A(z)z^4}\left(1-\frac{bz^4\phi'(z)^2}{r_{+}^2}\right)^{\frac{3}{2}}=0\\
    \label{eq36}
    &&\psi''(z)+\psi'(z)\left(\frac{A'(z)}{A(z)}-\frac{d-4}{z}\right) \nonumber \\
    &&+\frac{r_{+}^2}{A(z)z^4}\left(\frac{\phi^2(z)}{A(z)}-m^2\right)\psi(z)=0.
\end{eqnarray}
We can see that in the new coordinates the interval $r_{+}<r<\infty$ is now $1>z>0$, while the condition $\phi(r_{+})=0$ is now $\phi(1)=0$.
\section{\label{sec4} Critical temperature}
To obtain the critical temperature as a function of the charge density, we start with Eq.(\ref{eq35}). By definition, at the critical temperature we have that $\psi=0$. Therefore, Eq.(\ref{eq35}) becomes
\begin{eqnarray}
\label{eq37}
    \resizebox{0.42\textwidth}{!}{$\phi''(z)-\frac{\phi'(z)}{z}(d-4)+\frac{(d-2)bz^3}{r_{+(c)}^2}\phi'(z)^3=0,$}
\end{eqnarray}
where $r_{+(c)}$ is the horizon in this configuration.
We solve this equation as follows \cite{2012JHEP...05..002G,2016EPJC...76..146G}. Choosing that $b=0$, it reduces to
\begin{eqnarray}
    \label{eq38}
    \phi''(z)-\frac{\phi'(z)}{z}(d-4)=0.
\end{eqnarray}
Taking into account the boundary condition in Eq.(\ref{eq32}), the solution is:
\begin{eqnarray}
\label{eq39}
    \phi_{0}(z)=\lambda r_{+(c)}(1-z^{d-3}),
\end{eqnarray}
where
\begin{eqnarray}
\label{eq40}
    \lambda=\frac{\rho}{r_{+(c)}^{d-2}}.
\end{eqnarray}
After substituting this solution in the last term of Eq.(\ref{eq37}), we get
\begin{eqnarray}
    \label{eq41}
    &&\phi''(z)-\frac{\phi'(z)}{z}(d-4) \nonumber \\
    &&-b\lambda^3 r_{+(c)}(d-2)(d-3)^3z^{3(d-3)}=0.
\end{eqnarray}
Again using Eq.(\ref{eq32}), we obtain a solution to the first order of $b$:
\begin{eqnarray}
    \label{eq42}
    \resizebox{0.43\textwidth}{!}{$\phi(z)=\lambda r_{+(c)}\left((1-z^{d-3})-\frac{b\lambda_{0}^2(d-3)^3}{2(3d-7)}(1-z^{3d-7})\right),$}
\end{eqnarray}
where it has been used that $b\lambda^2=b\lambda_{0}^2+\mathcal{O}(b^2)$, and $\lambda_0^2$ is the value for $\lambda^2$ when $b=0$ \cite{2016EPJC...76..146G}.\\
If we express $A(z)$ to the first order of $ar_{0(c)}$ as
\begin{eqnarray}
    \label{eq43}
    A(z)=\frac{r_{+(c)}^2}{z^2}g(z),
\end{eqnarray}
with
\begin{eqnarray}
    \label{eq44}
    \resizebox{0.43\textwidth}{!}{$g(z)=\frac{1}{l^2}+\frac{d-1}{l^2}\left(-\frac{z^{d-1}}{d-1}+\frac{ar_{0(c)}z^{d-2}}{d-2}(1-z)\right),$}
\end{eqnarray}
for Eq.(\ref{eq36}) near $T_c$ we have
\begin{eqnarray}
    \label{eq45}
    &&\psi''(z)+\psi'(z)\left(\frac{g'(z)}{g(z)}-\frac{d-2}{z}\right) \nonumber \\
    &&+\frac{1}{g(z)}\left(\frac{\phi^2(z)}{g(z)r_{+(c)}^2}-\frac{m^2}{z^2}\right)\psi(z)=0.
\end{eqnarray}
We define near the AdS boundary \cite{2010JHEP...05..013S}:
\begin{eqnarray}
    \label{eq46}
    \psi(z)=\frac{\langle J \rangle}{r_{+(c)}^{\Delta_{+}}}z^{\Delta_{+}}F(z),
\end{eqnarray}
with $F(0)=1$ and $J$ being the condensation operator. Then, Eq.(\ref{eq45}) becomes \cite{2016EPJC...76..146G, 2018PhLB..781..139S}
\begin{eqnarray}
    \label{eq47}
    \begin{split}
    &F''(z)+\left(\frac{2\Delta_{+}-d+2}{z}+\frac{g'(z)}{g(z)}\right)F'(z) \\
    &+\bigg[\frac{\Delta_{+}(\Delta_{+}-1)}{z^2}+\left(\frac{g'(z)}{g(z)}-\frac{d-2}{z}\right)\frac{\Delta_{+}}{z} \\
    &-\frac{m^2}{g(z)z^2}\bigg]F(z)+\frac{\lambda^2}{g^2(z)}\bigg[(1-z^{d-3})^2 \\
    &-\frac{b\lambda_0^2(d-3)^3}{3d-7}(1-z^{d-3})(1-z^{3d-7})\bigg]F(z)=0,
    \end{split}
\end{eqnarray}
the solution of which has to satisfy the condition $F'(0)=0$. This equation can be transformed to the Sturm-Liouville form:
\begin{eqnarray}
    \label{eq48}
    \resizebox{0.42\textwidth}{!}{$\frac{d}{dz}(p(z)F'(z))+q(z)F(z)+\lambda^2r(z)F(z)=0,$}
\end{eqnarray}
where 
\begin{eqnarray}
    \label{eq49}
    &&p(z)=z^{2\Delta_{+}-d+2}g(z);\\
    \label{eq50}
    &&q(z)=z^{2\Delta_{+}-d+2}g(z)\bigg[\frac{\Delta_{+}(\Delta_{+}-1)}{z^2} \nonumber \\
    &&+\left(\frac{g'(z)}{g(z)}-\frac{d-2}{z}\right)\frac{\Delta_{+}}{z}-\frac{m^2}{g(z)z^2}\bigg];\\
    \label{eq51}
    &&r(z)=\frac{z^{2\Delta_{+}-d+2}}{g(z)}\bigg((1-z^{d-3})^2 \nonumber \\
    &&-\frac{b\lambda_{0}^2(d-3)^3}{3d-7}(1-z^{d-3})(1-z^{3d-7})\bigg).
\end{eqnarray}
Using the Sturm-Liouville eigenvalue problem, the eigenvalues of Eq.(\ref{eq48}) are
\begin{eqnarray}
    \label{eq52}
    \lambda^2=\frac{\int_{0}^{1}dz\left(p(z)F'(z)^2-q(z)F^2(z)\right)}{\int_{0}^{1}dz r(z)F^2(z)}.
\end{eqnarray}
We take the form of the trial function $F(z)$ to be $F(z)=1-\alpha z^2$, which obviously satisfies the boundary conditions. \\
From Eqs.(\ref{eq26}),(\ref{eq27}), and (\ref{eq40}) we get an important result for the dependence of $T_c$ on the charge density:
\begin{eqnarray}
    \label{eq53}
    T_c=\frac{d-1}{4\pi l^2\left(1-\frac{ar_{0(c)}}{d-2}\right)}\left(\frac{\rho}{\lambda}\right)^{\frac{1}{d-2}}.
\end{eqnarray}
From now on, we will take $m=-\frac{d-2}{l^2}$, which satisfies the Breitenlohner-Freedman (BF) bound \cite{BREITENLOHNER1982249,MEZINCESCU1985406,10.1007/978-3-319-12238-0_10}, and $l=1$ for simplicity. Therefore, we now have that $\Delta_{+}=d-2$. All this gives us the following expressions for Eqs.(\ref{eq49}), (\ref{eq50}), and (\ref{eq51}):
\begin{strip}
\begin{eqnarray}
    \label{eq54}
    &&p(z)=z^{d-2}\left[1+(d-1)\left(-\frac{z^{d-1}}{d-1}+\frac{ar_{0(c)}z^{d-2}(1-z)}{d-2}\right)\right]\\
    \label{eq55}
    &&q(z)=z^{d-2}\left[1+z^{d-2}(d-1)\left(\frac{ar_{0(c)}(1-z)}{d-2}-\frac{z}{d-1}\right)\right] \times \bigg\{ \frac{(d-2)(d-3)}{z^2}+\frac{d-2}{z}\bigg[\frac{(d-1)z^{d-3}ar_{0(c)}}{1-z^{d-1}} \nonumber \\
    &&-\frac{(d-1)z^{d-2}}{1-z^{d-1}} - \frac{(d-1)^2z^{d-2}ar_{0(c)}}{(d-2)(1-z^{d-1})}+\frac{z^{2d-4}(d-1)^2(1-z)ar_{0(c)}}{(d-2)(1-z^{d-1})^2}-\frac{d-2}{z}\bigg]+\frac{d-2}{z^2}\bigg[\frac{1}{1-z^{d-1}} \nonumber \\
    &&-\frac{ar_{0(c)}z^{d-2}(d-1)(1-z)}{(d-2)(1-z^{d-1})^2}\bigg] \bigg\}\\
    \label{eq56}
    &&r(z)=\frac{z^{d-2}(1-z^{d-3})}{1-z^{d-1}}\left(1-z^{d-3}-\frac{b\lambda_{0}^2(d-3)^3}{3d-7}(1-z^{3d-7})\right)\times\left(1-\frac{ar_{0(c)}z^{d-2}(d-1)(1-z)}{(1-z^{d-1})(d-2)}\right).
\end{eqnarray}
\end{strip}
\begin{figure}
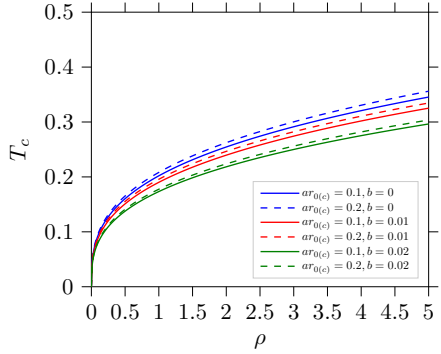
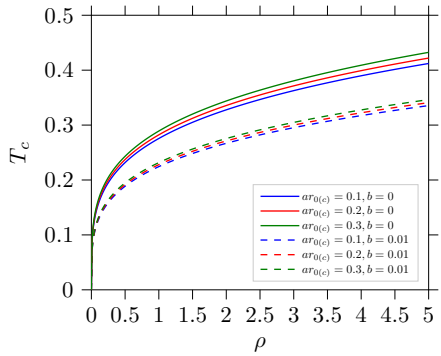
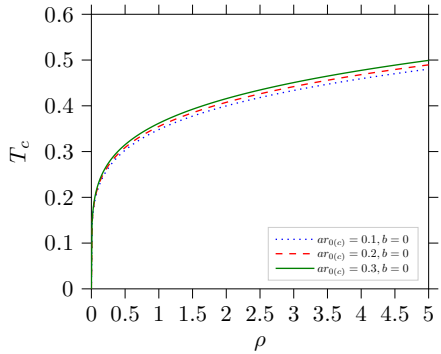

	\centering
	\begin{subfigure}[htb]{.36\textwidth}
		\centering
		\vfill

		\vfill
		\label{fig1:a}
		\caption{}
	\end{subfigure}
	\begin{subfigure}[htb]{.36\textwidth}
		\centering
		\vfill
		%
		\vfill
		\label{fig1:b}
		\caption{}
	\end{subfigure}
	\begin{subfigure}[htb]{.36\textwidth}
		\raggedleft
		\vfill
		%
		\vfill
		\label{fig1:c}
		\caption{}
	\end{subfigure}
	\caption{\label{fig1}$T_c$ as a function of $\rho$ for $d=5$ (a), $d=6$ (b), $d=7$ (c).}
\end{figure}
Here, we will analyze the cases for $d=5, 6, 7$. We proceed by calculating $\lambda_0$ for $b=0$ from Eq.(\ref{eq52}). Then, we find the value of $\alpha$ for which $\lambda_0$ is at its minimum. After that we repeat this procedure for $b=0.01$ and $b=0.02$. As noted earlier, we focus on the quantity $ar_{0(c)}$, which is varied to get different results. This makes the calculations much simpler and more accurate than if we vary, for example, $a\rho^{\frac{1}{d-2}}$, and leads to the same findings. The results are shown in Tables \ref{tab1}, \ref{tab2}, and \ref{tab3}. It is interesting to note that for $d=6$ and $d=7$ there are no meaningful values for $b=0.02$ and $b=0.01$, $b=0.02$, respectively. The reason for this is that for each $d$ and for different $ar_{0(c)}$, there is a limiting value for $b$, above which we cannot get algebraically reasonable results, as $\lambda^2$ is negative. Table \ref{tab4} shows these limiting values for the different configurations. Thus, it can be confirmed that the approximation for small $b$ is accurate. \\
Some plots showing $T_c$ as a function of $\rho$ are also shown (Fig. \ref{fig1}). As expected, the critical temperature decreases with increasing $b$. We can see that for the configuration that was chosen, $T_c$ increases for larger $ar_{0(c)}$. Of course, at first glance this itself does not mean that, for a given $\rho$, $a$ is increasing with $ar_{0(c)}$. However, one can directly
check using Eqs.(\ref{eq26}), (\ref{eq40}) and the values from Tables \ref{tab1}, \ref{tab2}, \ref{tab3} that for increasing $ar_{0(c)}$ and given $b$ and $\rho$, both values of $a$ and $r_{0(c)}$ rise (the 5th and 6th columns of the same tables). Therefore, as we choose some $\rho$, there is an increase in $a$ when $ar_{0(c)}$ is larger. Thus, the plots show that the critical temperature rises as $a$ increases, meaning that the condensation is easier to form.
\begingroup
\begin{table}[t]
    \centering
    \renewcommand{\arraystretch}{1.3}
    \resizebox{\columnwidth}{!}{
    \begin{tabular}{ccccccc}
         $ar_{0(c)}$&$b$&$\alpha$&$\lambda^2$&$(a\rho^{1/3})$&$(r_{0(c)}/\rho^{1/3})$&$(T_c/\rho^{1/3})$\\ \hline
         \multirow{3}{*}{0.1}&0&0.7285&18.76&0.1576&0.6346&0.2020 \\
         &0.01&0.7608&26.95&0.1674&0.5974&0.1902 \\
         &0.02&0.8301&47.01&0.1836&0.5445&0.1733 \\
         \hline
         \multirow{3}{*}{0.2}&0&0.7349&19.31&0.3057&0.6541&0.2082 \\
         &0.01&0.7674&28.04&0.3254&0.6147&0.1957 \\
         &0.02&0.8391&50.22&0.3585&0.5578&0.1776 \\
         \hline
         \multirow{3}{*}{0.3}&0&0.7412&19.86&0.4443&0.6752&0.2149 \\
         &0.01&0.7737&29.17&0.4737&0.6333&0.2016 \\
         &0.02&0.8480&53.74&0.5245&0.5720&0.1821\\
    \end{tabular}
    }
    \caption{Values of the parameters for $d=5$}
    \label{tab1}
\end{table}
\endgroup
\begingroup
\begin{table}[t]
    \centering
    \renewcommand{\arraystretch}{1.3}
    \resizebox{\columnwidth}{!}{
    \begin{tabular}{ccccccc}
         $ar_{0(c)}$&$b$&$\alpha$&$\lambda^2$&$(a\rho^{1/4})$&$(r_{0(c)}/\rho^{1/4})$&$(T_c/\rho^{1/4})$\\ \hline
         \multirow{3}{*}{0.1}&0&0.7963&23.09&0.1444&0.6927&0.2756 \\
         &0.01&0.9856&119.8&0.1773&0.5639&0.2244 \\
         &0.02&-&-&-&-&- \\
         \hline
         \multirow{3}{*}{0.2}&0&0.8000&23.53&0.2820&0.7093&0.2822 \\
         &0.01&0.9950&129.9&0.3491&0.5729&0.2279 \\
         &0.02&-&-&-&-&- \\
         \hline
         \multirow{3}{*}{0.3}&0&0.8036&23.96&0.4128&0.7268&0.2892 \\
         &0.01&1.005&141.4&0.5153&0.5822&0.2316 \\
         &0.02&-&-&-&-&-\\
    \end{tabular}
    }
    \caption{Values of the parameters for $d=6$}
    \label{tab2}
\end{table}
\endgroup
\begingroup
\begin{table}[htb]
    \centering
    \renewcommand{\arraystretch}{1.3}
    \resizebox{\columnwidth}{!}{
    \begin{tabular}{ccccccc}
         $ar_{0(c)}$&$b$&$\alpha$&$\lambda^2$&$(a\rho^{1/5})$&$(r_{0(c)}/\rho^{1/5})$&$(T_c/\rho^{1/5})$\\ \hline
         \multirow{3}{*}{0.1}&0&0.8401&28.92&0.1372&0.7289&0.3480 \\
         &0.01&-&-&-&-&- \\
         &0.02&-&-&-&-&- \\
         \hline
         \multirow{3}{*}{0.2}&0&0.8425&29.31&0.2692&0.7431&0.3548 \\
         &0.01&-&-&-&-&- \\
         &0.02&-&-&-&-&- \\
         \hline
         \multirow{3}{*}{0.3}&0&0.8447&29.7&0.3958&0.7579&0.3619 \\
         &0.01&-&-&-&-&- \\
         &0.02&-&-&-&-&-\\
    \end{tabular}
    }
    \caption{Values of the parameters for $d=7$}
    \label{tab3}
\end{table}
\endgroup
\begingroup
\begin{table}[htb]
    \centering
    \renewcommand{\arraystretch}{1.3}
    \resizebox{\columnwidth}{!}{
    \begin{tabular}{cccc}
         &$ar_{0(c)}=0.1$&$ar_{0(c)}=0.2$&$ar_{0(c)}=0.3$\\ \hline
         $b\text{ }(d=5)$&0.0385&0.0374&0.0364 \\
         $b\text{ }(d=6)$&0.0138&0.0135&0.0133 \\
         $b\text{ }(d=7)$&0.00615&0.00607&0.00599 \\
    \end{tabular}
    }
    \caption{Limiting values for $b$}
    \label{tab4}
\end{table}
\endgroup
\section{\label{sec5} The condensate}
We continue by analyzing how the Born-Infeld parameter and the $f(R)$ gravity corrections affect the values of the condensation operator near $T_c$. The procedure in \cite{2016EPJC...76..146G} will be followed (see also \cite{2012JHEP...05..002G, 2012JHEP...05..156G, 2018PhLB..781..139S}). To that end, we use Eqs.(\ref{eq35}) and (\ref{eq46}) to get:
\begin{eqnarray}
    \label{eq57}
    &&\phi''(z)-\frac{\phi'(z)}{z}(d-4) \nonumber \\
    &&+\frac{(d-2)bz^3\phi'(z)^3}{r_{+}^2}=\frac{\langle J \rangle^2}{r_{+}^2}\mathcal{B}(z)\phi(z)
\end{eqnarray}
with $\mathcal{B}(z)=\frac{2z^{2\Delta_{+}-4}F^2(z)}{r_{+}^{2\Delta_{+}-4}A(z)}\left(1-\frac{bz^4\phi'(z)^2}{r_{+}^2}\right)^{\frac{3}{2}}$. After expanding $\phi(z)$ including the small term $\frac{\langle J \rangle^2}{r_{+}^2}$, we have:
\begin{eqnarray}
    \label{eq58}
    \frac{\phi(z)}{r_{+}}=&&\lambda \left((1-z^{d-3})-\frac{b\lambda_{0}^2(d-3)^3}{2(3d-7)}(1-z^{3d-7})\right) \nonumber \\
    &&+\frac{\langle J \rangle^2}{r_{+}^2}\zeta(z),
\end{eqnarray}
where $\zeta(1)=\zeta'(1)=0$. As we substitute this equation in Eq.(\ref{eq57}) and compare the coefficients of $\frac{\langle J \rangle^2}{r_{+}^2}$, we obtain:
\begin{eqnarray}
    \label{eq59}
    \zeta''(z)&&-\left(\frac{d-4}{z}+3b\lambda_{0}^2(d-2)(d-3)^2z^{2d-5}\right)\zeta'(z) \nonumber \\
    &&=\lambda \frac{2z^{2\Delta_{+}-4}F^2(z)}{r_{+}^{2\Delta_{+}-4}A(z)}\mathcal{A}_1(z),
\end{eqnarray}
with 
\begin{eqnarray}
\label{eq60}
\mathcal{A}_{1}(z)=&&1-z^{d-3}-\frac{3b\lambda_{0}^2(d-3)^2}{2}\bigg((1-z^{d-3})z^{2d-4} \nonumber \\
&&+\frac{d-3}{3(3d-7)}(1-z^{3d-7})\bigg).
\end{eqnarray}
We solve this equation by multiplying it by $z^{-(d-4)}e^{\frac{3(d-2)(d-3)^2b\lambda_0^{2}z^{2d-4}}{2d-4}}$, which gives
\begin{strip}
\begin{eqnarray}
    \label{eq61}
    \frac{d}{dz}\left(z^{-(d-4)}e^{\frac{3(d-2)(d-3)^2b\lambda_0^{2}z^{2d-4}}{2d-4}} \zeta'(z)\right)=\lambda \frac{2z^{d-2}F^2(z)}{r_{+}^{2d-6}g(z)}e^{\frac{3(d-2)(d-3)^2b\lambda_0^{2}z^{2d-4}}{2d-4}}\mathcal{A}_{1}(z),
\end{eqnarray}
\end{strip}
where we used Eq.(\ref{eq43}). After integration in the range $(0,1)$, taking into account the boundary conditions stated above, we arrive at the following result:
\begin{eqnarray}
    \label{eq62}
    \frac{\zeta'(z)}{z^{d-4}}\bigg\rvert_{z\to 0}=-\frac{\lambda}{r_{+}^{2d-6}}\mathcal{A}_2,
\end{eqnarray}
with
\begin{eqnarray}
    \label{eq63}
    \resizebox{0.43\textwidth}{!}{$\mathcal{A}_2=\int_{0}^{1}dz\frac{2z^{d-2}F^2(z)}{g(z)}e^{\frac{3(d-2)(d-3)^2b\lambda_0^{2}z^{2d-4}}{2d-4}}\mathcal{A}_1(z).$}
\end{eqnarray}
Now, we will look at the asymptotic behaviour of $\phi(z)$. We already know that it satisfies Eq.(\ref{eq32}). If we compare this equation with Eq.(\ref{eq58}) in this limit, we get
\begin{eqnarray}
    \label{eq64}
    &&\mu - \frac{\rho}{r_{+}^{d-3}}z^{d-3}=\lambda r_{+}\bigg[(1-z^{d-3}) \nonumber \\
    &&-\frac{b\lambda_{0}^2(d-3)^3}{2(3d-7)}(1-z^{3d-7})\bigg]+\frac{\langle J \rangle^2}{r_{+}}\bigg[\zeta(0) \nonumber \\
    &&+z\zeta'(0)+\ldots +\frac{\zeta^{(d-3)}(0)}{(d-3)!}z^{d-3}+\ldots\bigg],
\end{eqnarray}
where $\zeta^{(d-3)}(0)$ denotes the derivative of order $(d-3)$ at $z=0$. If we compare the coefficients of $z^{d-3}$, we see that
\begin{eqnarray}
    \label{eq65}
    -\frac{\rho}{r_{+}^{d-3}}=-\lambda r_{+}+\frac{\langle J \rangle^2 \zeta^{(d-3)}(0)}{r_{+}(d-3)!}.
\end{eqnarray}
Thus, if we require that
\begin{eqnarray}
    \label{eq66}
    \frac{\zeta^{(d-3)}(0)}{(d-4)!}=\frac{\zeta'(z)}{z^{d-4}}\bigg\rvert_{z\to 0},
\end{eqnarray}
and using  Eqs.(\ref{eq62}) and (\ref{eq65}), we obtain:
\begin{eqnarray}
\label{eq67}
    \frac{\rho}{r_{+}^{d-2}}=\lambda\left(1+\frac{\langle J \rangle^2 \mathcal{A}_2}{r_{+}^{2d-4}(d-3)}\right).
\end{eqnarray}
Taking into account Eqs.(\ref{eq40}), (\ref{eq53}) and that $T\to T_c$, this reduces to
\begin{eqnarray}
    \label{eq68}
    \langle J \rangle^2=&&\frac{(d-3)}{\mathcal{A}_2}\left(\frac{4\pi T_c\left(1-\frac{ar_{0(c)}}{d-2}\right)}{d-1}\right)^{2d-4} \nonumber \\
    &&\times \left(1-\left(\frac{T}{T_c}\right)^{d-2}\right).
\end{eqnarray}
\begingroup
\begin{table}[t]
    \centering
    \renewcommand{\arraystretch}{1.3}
    \resizebox{\columnwidth}{!}{
\begin{tabular}{cccccc}
         $ar_{0(c)}$&$b$&$\alpha$&$\lambda^2$&$\beta$\\ \hline
         \multirow{3}{*}{0.1}&0&0.7285&18.76&218.973 \\
         &0.01&0.7608&26.95&248.817 \\
         &0.02&0.8301&47.01&296.439 \\
         \hline
         \multirow{3}{*}{0.2}&0&0.7349&19.31&199.961 \\
         &0.01&0.7674&28.04&227.746 \\
         &0.02&0.8391&50.22&272.631 \\
         \hline
         \multirow{3}{*}{0.3}&0&0.7412&19.86&181.881 \\
         &0.01&0.7737&29.17&207.645 \\
         &0.02&0.8480&53.74&249.793\\
    \end{tabular}
    }
    \caption{Values of $\beta$ for $d=5$}
    \label{tab5}
\end{table}
\endgroup
\begingroup
\begin{table}[htb]
    \centering
    \renewcommand{\arraystretch}{1.3}
    \resizebox{\columnwidth}{!}{
    \begin{tabular}{cccccc}
         $ar_{0(c)}$&$b$&$\alpha$&$\lambda^2$&$\beta$\\ \hline
         \multirow{3}{*}{0.1}&0&0.7963&23.09&488.402 \\
         &0.01&0.9856&119.8&803.154 \\
         &0.02&-&-&- \\
         \hline
         \multirow{3}{*}{0.2}&0&0.8000&23.53&444.717 \\
         &0.01&0.9950&129.9&738.036 \\
         &0.02&-&-&- \\
         \hline
         \multirow{3}{*}{0.3}&0&0.8036&23.96&403.792 \\
         &0.01&1.005&141.4&676.475 \\
         &0.02&-&-&-\\
    \end{tabular}
    }
    \caption{Values of $\beta$ for $d=6$}
    \label{tab6}
\end{table}
\endgroup
\begingroup
\begin{table}[htb]
    \centering
    \renewcommand{\arraystretch}{1.3}
    \resizebox{\columnwidth}{!}{
    \begin{tabular}{cccccc}
         $ar_{0(c)}$&$b$&$\alpha$&$\lambda^2$&$\beta$\\ \hline
         \multirow{3}{*}{0.1}&0&0.8401&28.92&772.493 \\
         &0.01&-&-&- \\
         &0.02&-&-&- \\
         \hline
         \multirow{3}{*}{0.2}&0&0.8425&29.31&702.237 \\
         &0.01&-&-&- \\
         &0.02&-&-&- \\
         \hline
         \multirow{3}{*}{0.3}&0&0.8447&29.7&636.944 \\
         &0.01&-&-&- \\
         &0.02&-&-&-\\
    \end{tabular}
    }
    \caption{Values of $\beta$ for $d=7$}
    \label{tab7}
\end{table}
\endgroup
Therefore,
\begin{eqnarray}
    \label{eq69}
    \langle J \rangle = \beta T_{c}^{d-2}\sqrt{1-\frac{T}{T_c}},
\end{eqnarray}
where
\begin{eqnarray}
    \label{eq70}
    \resizebox{0.42\textwidth}{!}{$\beta=\sqrt{\frac{(d-3)(d-2)}{\mathcal{A}_2}}\left(\frac{4\pi \left(1-\frac{ar_{0(c)}}{d-2}\right)}{d-1}\right)^{d-2}.$}
\end{eqnarray}
\begin{figure}
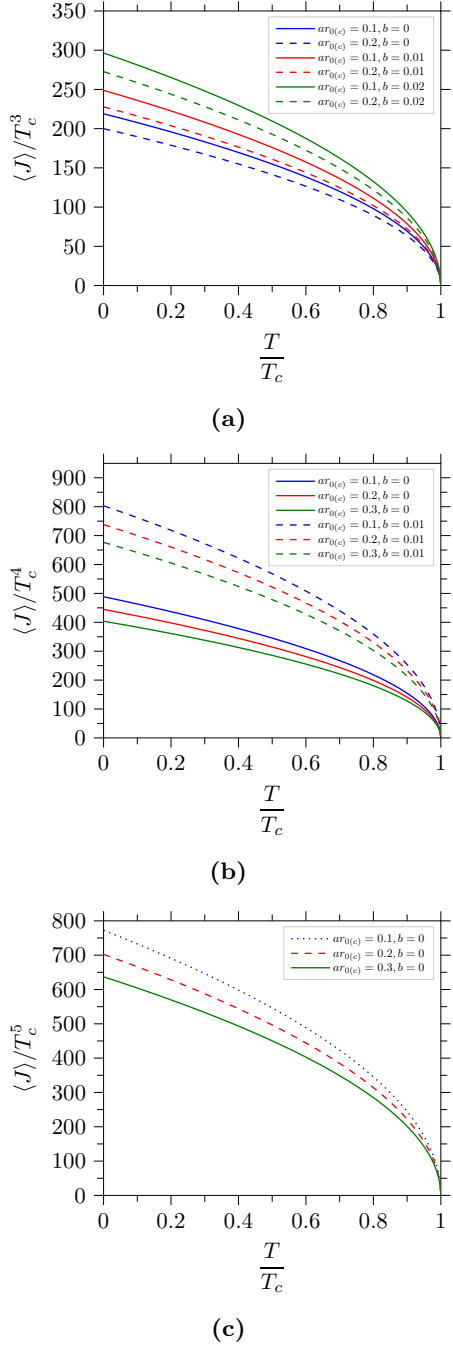

	\centering
	\begin{subfigure}[t]{.45\textwidth}
		\centering
		\vfill
		%
		\vfill
		\label{fig2:a}
		\caption{}
	\end{subfigure}
	\begin{subfigure}[t]{.45\textwidth}
		\centering
		\vfill
		%
		\vfill
		\label{fig2:b}
		\caption{}
	\end{subfigure}
	\begin{subfigure}[t]{.45\textwidth}
		\centering
		\vfill
		%
		\vfill
		\label{fig2:c}
		\caption{}
	\end{subfigure}
	\caption{\label{fig2}$\frac{\langle J \rangle}{T_c^{d-2}}$ as a function of $\frac{T}{T_c}$ for $d=5$ (a), $d=6$ (b), $d=7$ (c).}
\end{figure}
As expected, the critical exponent is $1/2$. We now calculate $\mathcal{A}_1$, $\mathcal{A}_2$ and $\beta$ for $d=5,6$ and $7$ (a similar procedure for $d=5$ was shown in \cite{2016EPJC...76..146G}). The results are shown in Tables \ref{tab5}, \ref{tab6} and \ref{tab7}. Plots of $\frac{\langle J \rangle}{T_c^{d-2}}$ as a function of $\frac{T}{T_c}$ are also shown in Fig. \ref{fig2}. The condensation gap increases with larger values of $b$. One can also see that it decreases with increasing $ar_{0(c)}$ and, hence, with increasing $a$, as noted earlier.
\section{\label{sec6} Conclusion}
In this paper, higher dimensional holographic superconductors were analyzed in the background of modified $f(R)$ gravity and Born-Infeld electrodynamics using the Sturm-Liouville method. For the model of $f(R)$ gravity that was used, a small correction parameter was introduced, which allows a perturbative analysis of the problem. The calculations of the critical temperature and the dimensionless condensation show that the higher values of the Born-Infeld parameter make the condensation harder to form. On the other hand, increasing $f(R)$ gravity modifications make the formation of the condensate easier. These results were shown in $d=5,6$ and $7$ dimensions. In addition, the limiting values of the Born-Infeld parameter, above which one cannot get meaningful results, were found. This gives the range of applicability of Born-Infeld electrodynamics in the configuration that is considered. It is important to note that the analytical methods in this paper have been thoroughly investigated in other cases, and were shown to give a very good accuracy. In particular, the Sturm-Liouville method has been known to agree very well with numerical results, while also being more accurate than the matching method. \\
It would be extremely interesting if the results in this paper can be extended away from the probe limit in future work, or if other models of $f(R)$ gravity can be similarly analyzed.
\section*{\small Acknowledgements \normalfont
I would like to thank Ivelina Yovkova, Dragomir Roussev and Milena Boeva for all their love and support.}
\section*{\small Data Availability Statement \normalfont This manuscript has no associated data or the data will not be deposited. [Authors' comment: This is a theoretical study. No data to deposit. Figures can be computed using the equations in the paper.]}

\end{document}